\documentstyle[preprint,aps,epsfig]{revtex}

\newcommand{\bea}{\begin{eqnarray}}
\newcommand{\eea}{\end{eqnarray}}
\newcommand{\be}{\begin{equation}}
\newcommand{\ee}{\end{equation}}

\begin{document}


\preprint{
\hspace{-30.5mm}
\raisebox{2.8ex}{SNUTP 99--047}
}

\title{Spin configuration of top quark pair production
\\
with large extra dimensions at photon-photon colliders 
}

\author{
Kang Young Lee$^a$,
Seong Chan Park$^b$,
H. S. Song$^{a,b}$,
\\
JeongHyeon Song$^a$,
and Chaehyun Yu$^b$
}
\vspace{1.0cm}

\address{
$^a$Center for Theoretical Physics, Seoul National University,
Seoul 151--742, Korea\\
$^b$Department of Physics, Seoul National University, Seoul 151--742, Korea
}

\maketitle


\begin{abstract}

Top quark pair production at photon-photon colliders is studied 
in low scale quantum gravity scenario. From the dependence of 
the cross sections on the spin configuration of the top quark 
and anti-quark, we introduce a new observable, top spin asymmetry.
It is shown that there exists a special top spin basis where 
with the polarized parent electron beams the top spin asymmetry 
vanishes in the standard model but retains substantial values 
with the large extra dimension effects. We also present lower 
bounds of the quantum gravity scale $M_S$ from total cross 
sections with various combinations of the laser, electron beam, 
and top quark pair polarizations. The measurements of the top 
spin state $(t_\uparrow\bar{t}_\downarrow)$ with unpolarized 
initial beams are shown to be most effective, enhancing by about 
5\% the $M_S$ bounds with respect to totally unpolarized case.
\end{abstract}

\pacs{ }

\narrowtext


\section{Introduction}

Conventional wisdom is that gravitational interactions
suppressed by the Planck scale $M_{Pl}$ 
are to be neglected at any collider.
Recently Arkani-Hamed, Dimopoulos, and Dvali 
suggested that quantum gravity may become strong at the
electroweak scale by introducing large extra dimensions 
and by confining the standard model (SM) fields to a $(3+1)$-dimensional 
hypersurface\cite{ADD}.
The observed weakness of macroscopic gravity can be attributed
to the largeness of the compactified extra dimensions.
Since the fundamental scale $M_S$
can then be around TeV scale, the hierarchy
between the electroweak scale and the Planck scale is relieved.

Extra dimension effects are recast 
in an effective theory valid below the $M_S$,
as a massless graviton propagating 
in the full $(4+n)$-dimensional space is replaced by massive
Kaluza-Klein (KK) gravitons
propagating in four-dimensional subspace\cite{Lykken}.
Since the large phase space of the KK modes 
enhances the strength of their couplings to the SM particles
from $\sim 1/M_{Pl}$ into $\sim 1/M_S$,
high energy colliders may provide some distinctive experimental
signatures, questioning the above wisdom.
Two kinds of collider signals have been studied. 
Direct signals contain the production of
KK states, spin-2 neutral particles\cite{direct}.
Since they escape collider detectors,
the signal is to be recognized
as missing energy.
In general, the production of KK gravitons is quite sensitive to the 
number of extra dimensions.
Indirect signals with intermediate KK states have less 
sensitivity to the number of extra dimensions,
which are preferable to obtain experimental constraints on
the scale $M_S$\cite{indirect}.  
Moreover spin-2 nature of the KK modes 
leaves characteristic spin configuration of 
outgoing particles.

In particular, top pair production holds an advantageous position
to draw the spin information of the process.
Very heavy mass of top quark \cite{TOP} prompts
itself to decay before hadronization,
leaving alive the top spin information which can be determined
from the angular distribution of its electroweak decay products\cite{SPIN}.
Thus the top spin configuration can provide additional observables
of new physics, particularly when high spin particles such as gravitons
are involved.
At $e^+ e^-$ linear colliders (LC),
it has been shown that the $s$-channel diagram mediated by
spin-2 KK gravitons significantly modifies 
the pattern of spin configuration of the 
top quark and top anti-quark\cite{top-LSSY}.

Photon-photon collisions have been regarded 
as one of the best alternatives of $e^+ e^-$ collisions at LC,
where high energy photon beams can be achieved through laser 
back-scattering of the parent $e^+e^-$ beams.
The controllability of the laser and electron beam polarizations
provides good opportunities to probe new physics.
It is shown that
photon colliders are sensitive to the presence of large extra 
dimensions, yielding higher low bound of the $M_S$
than any other collider\cite{gg}.
The role of polarizations, however,
has been studied only about the laser and parent electron beams.

In this paper we study the large extra dimension effects on the
polarizations of top quark pair production at photon colliders,
including the top spin configuration
which shall be shown to provide a unique channel.
From the total cross sections in various top spin bases,
we will observe that there is no particular top spin basis 
where the pattern of the spin configuration
is crucially modified
by the large extra dimension effects.
Instead a new observable,
the `top spin asymmetry', is to be introduced.
There exists a special spin basis, in particular with polarized
electron beams,
where this top spin asymmetry vanishes in the SM but has
substantial values with low scale quantum gravity effects.

This paper is organized as follows.
In Sec.~II we present the scattering amplitudes of the 
$\gamma\gamma \rightarrow t \bar{t}$ process
corresponding to the polarizations of initial photons
in a general spin basis of the top quark 
and top anti-quark.
The differences from the process $e^+e^-\rightarrow t \bar{t}$ are 
also discussed.
Since photon beams with high energy and high luminosity
are to be obtained from $e^+ e^-$ LC,
we review practical cross sections by taking into account of 
backward Compton scattering of laser beams from high energy $e^+ e^-$ beams.
Section III is devoted to the results and discussions.
The definition and numerical results of the top spin asymmetry 
along with an optimal top spin basis shall be given.
We also present at 2$\sigma$ level the lower bound of $M_S$ 
which top spin information can enhance by a few percents.
Finally Sec.~IV deals with our conclusions.


\section{Formalism}

For the process
\begin{equation}
\label{process}
\gamma(k_1,\lambda_1) + \gamma(k_2,\lambda_2)
\rightarrow t(p_1,\kappa_1) + \bar{t}(p_2, \kappa_2),
\end{equation}
there are two types of tree level diagrams in the SM,
$t$-channel and $u$-channel ones.
In Eq.~(\ref{process}), $\lambda_i$ and $\kappa_i$ $(i=1,2)$
denote the polarizations of photons and top quarks, respectively.
The effective theory of low scale quantum gravity allows an
$s$-channel diagram mediated by virtual KK graviton modes (see Fig.~1),
of which the scattering amplitude is
\begin{equation}
\label{amp}
{\cal M}_G = - \frac{2s\lambda}{M_S^4}\,\overline{u}(p_1)
   \left[ ( j\cdot \varepsilon_2) \rlap/{\varepsilon}_1
   +( j\cdot \varepsilon_1) \rlap/{\varepsilon}_2
   + 2 {\varepsilon}_1 \cdot {\varepsilon}_2
   \left(
	\frac{(j\cdot k_2) \rlap/k_1 + (j\cdot k_1) \rlap/k_2 }{s}-m_t
	\right)
\right] v(p_2)\,,
\end{equation}
where $j=p_1-p_2$,
and $\varepsilon_i=\varepsilon(k_i)\,\,(i=1,2)$ are 
the polarization vectors of initial photons. 
All the ambiguity such as the number of extra dimensions
and the compactification models is encoded by order one parameter $\lambda$.
Hereafter $\lambda=\pm1$ cases are to be considered,
which are sufficient for an
estimation of the scale $M_S$.

Let us review a general spin basis for top quark 
and top anti-quark \cite{generic,Hori}.
Since $CP$ invariance holds in the graviton-mediated diagram as well
as in the SM, the spins of the top quark pair lie
in the production plane.
First defining the spin states of the
top quark and anti-quark in their own rest frame,
we decompose, respectively, their spins along the reference axes
$\hat{\eta}$ and $\hat{\bar{\eta}}$
which are chosen back to back in the zero momentum frame.
Thus only one parameter $\xi$ can specify 
both $\hat{\eta}$ and $\hat{\bar{\eta}}$;
the $\xi$ is defined in the top quark rest frame
as the clockwise angle between the three-momentum 
of top anti-quark and the up-state spin 
of top quark, as depicted in Fig.~2.
The usual helicity basis is obtained by setting $\xi=0$ or $\pi$.

When the top quark is emitted in the $z$-axis,
the polarization vectors of initial photons 
are given by, 
in the center of momentum (CM) frame with the Coulomb gauge,
\begin{equation}
\label{pol}
\varepsilon_1(\pm)=\varepsilon_2(\mp)=\frac{1}{\sqrt{2}}
\big( 0, \mp \cos\theta, -i, \mp\sin\theta\big),
\end{equation}
where $\theta$ is the scattering angle.
There is a relation\cite{CCIS}, usually being used in the squared amplitudes,
\bea
\label{ggpol}
\varepsilon^i(k,\lambda_1) \varepsilon^{*j}(k,\lambda_2)
&=&
\frac{1}{2}
\left[
(\delta^{i j} - \hat{k}^i  \hat{k}^j ) \delta_{\lambda_1\lambda_2}
-\frac{i}{2} (\lambda_1 +\lambda_2)\epsilon^{ijk} \hat{k}^k \right.
\\ \nonumber &&\quad
-\frac{1}{2}(\lambda_1 -\lambda_2)
\left\{
\hat{a}^i(\hat{\bf k} \times \hat{\bf a})^j 
+ \hat{a}^j(\hat{\bf k} \times \hat{\bf a})^i
\right\}
\\ \nonumber &&\quad
\left.
+\frac{1}{2}(\lambda_1\lambda_2-1)
\left\{
\hat{a}^i\hat{a}^j
-(\hat{\bf k} \times \hat{\bf a})^i (\hat{\bf k} \times \hat{\bf a})^j
\right\}
\right]
\,,
\eea
where $\hat{\bf a}$ is an arbitrary unit vector perpendicular to $\hat{\bf k}$,
and $i,j=1,2,3$.

Equation (\ref{ggpol}) can be applied even in the amplitude level.
In the process of Eq.~(\ref{process}),
two channels are possible according to the total angular momentum:
$J_z = 0 $ and $J_z = 2 $ states. 
In the $J_z =0$ states ($\gamma_R\gamma_R$ or $\gamma_L\gamma_L$),
two polarization vectors are related by
$\varepsilon_2 (+)= -\varepsilon_1 ^\ast (+)$,
as can be easily seen from Eq.~(\ref{pol}).
Then the $\varepsilon^\mu_1\varepsilon^\nu_2$
in Eq.~(\ref{amp}) can be replaced by
\be
\label{polpol}
\varepsilon^i_1(+1)
\varepsilon^j_2(+1)
=
-\varepsilon^i(k_1,+1) \varepsilon^{*j}(k_1,+1)
\,.
\ee
Note that only the first term in Eq.~(\ref{ggpol}) 
contributes to the scattering amplitude in Eq.~(\ref{amp})
since the last two terms vanish in the case of $\lambda_1=\lambda_2$,
and the anti-symmetric second term does not contribute to 
the amplitude symmetric under the exchange of two photons.
In the CM frame where $ j \cdot k_1=- j \cdot k_2 =-{\bf j}\cdot{\bf k_1}$
and $\bar{u}(p_1) \rlap/k_1 v(p_2) = -\bar{u}(p_1) \rlap/k_2 v(p_2) 
=\bar{u}(p_1) \gamma_i \, v(p_2) k_1^i$,
the scattering amplitude mediated by the KK modes
vanishes:
\be
\label{zero-amp}
{\mathcal M}^{RR}_G
=
\frac{2s\lambda}{M_S^4}
\left[
\bar{u}\gamma_i v j^j \left(
\delta^{ij}-\frac{k_1^i k_1^j}{k_1^0 k_1^0 } \right)
-\bar{u}\gamma_i v j^i 
+\frac{ \bar{u}\gamma_i v k_1^i ({\bf j}\cdot{\bf k}_1)}{k_1^0 k_1^0}
\right]
 =0
\,.
\ee
Thus the effects of low scale quantum gravity exist only in
the $J_z=2$ channels. 

The squared amplitudes for each top spin 
configuration with the large extra dimension effects are,
with $\beta=\sqrt{1-4m_t^2/s}$,
\begin{eqnarray}
\label{ampsq}
|{\cal M}|^2 (\gamma_R \gamma_L \rightarrow
t_\uparrow \overline{t}_\uparrow ~{\rm or}~ t_\downarrow
\overline{t}_\downarrow ) &=& \frac{1}{4}N_cs^2
\beta^2\sin^2\theta[D_t+D_u+D_s]^2F_1^2,
\nonumber \\
|{\cal M}|^2  (\gamma_R \gamma_L \rightarrow
t_\uparrow \overline{t}_\downarrow ~{\rm or}~ t_\downarrow
\overline{t}_\uparrow ) &=& \frac{1}{4}N_cs^2
\beta^2\sin^2\theta[D_t+D_u+D_s]^2(F_2\mp1)^2,
\nonumber \\ 
|{\cal M}|^2  (\gamma_R \gamma_R \rightarrow
t_\uparrow \overline{t}_\uparrow ~{\rm or}~ t_\downarrow
\overline{t}_\downarrow ) &=& \frac{1}{4}N_cs^2
(1-\beta^2)[D_t+D_u]^2(1\mp\beta\cos\xi)^2,
\nonumber \\
|{\cal M}|^2(\gamma_R \gamma_R \rightarrow
t_\uparrow \overline{t}_\downarrow ~{\rm or}~ t_\downarrow
\overline{t}_\uparrow ) &=& \frac{1}{4}N_cs^2
\beta^2(1-\beta^2)[D_t+D_u]^2\sin^2\xi,
\end{eqnarray}
where $N_c$ is the number of color, 
$D_{s,t,u}$ are the effective propagation factors defined by 
\begin{equation}
D_t=\frac{(Q_te)^2}{t-m_t^2},\quad
D_u=\frac{(Q_te)^2}{u-m_t^2},\quad
D_s=\frac{4s\lambda}{M_S^4},
\end{equation}
and $F_{1,2}$ are the spin configuration factors as
functions of the scattering angle, given by
\begin{eqnarray}
F_1&=&\sqrt{1-\beta^2}\sin\theta\cos\xi-\cos\theta\sin\xi, 
\nonumber \\
F_2&=&\sqrt{1-\beta^2}\sin\theta\sin\xi+\cos\theta\cos\xi.
\end{eqnarray}
The SM results are in agreement with those of Ref.~\cite{Hori}.
It has been noted that the processes
$\gamma_R\gamma_L \to (t_\uparrow \bar{t}
_\uparrow ~{\rm or}~ t_\downarrow\bar{t}_\downarrow)$
permit a special top spin
basis where the SM prediction vanishes.
The other amplitudes can be obtained by using the
$CP$ invariance such as 
\begin{equation}
\label{cp}
|{\cal M}_{RL\uparrow \uparrow}| = |{\cal M}_{LR\downarrow \downarrow}|,
\quad
|{\cal M}_{RR\uparrow \downarrow}| = |{\cal M}_{LL\downarrow \uparrow}|
,
\end{equation}
where, e.g., the suffix $RL\uparrow\uparrow$ denotes the process
$\gamma_R\gamma_L\rightarrow t_\uparrow \bar{t}_\uparrow$. 

The squared amplitudes in Eq.~(\ref{ampsq})
imply important characteristic features of the $\gamma\gamma \to t \bar{t}$
process.
First the low scale quantum gravity effects contained in $D_s$
exist only in the $J_z=2$ cases,
as discussed before.
Second the effects in any top spin basis do not modify
the top spin configuration.
This is not the case at $e^+e^-$ colliders:
for instance
the quantum gravity leads to non-zero observables for like-spin
states of the top quark pair in a special
spin basis where the SM prediction vanishes\cite{top-LSSY}.

In order to draw the information of large extra dimensions from 
the top spin configuration,
we take notice of the following points:
the quantum gravity effects, 
even being factored out from the spin configuration,
appear in the $J_z=2$ cases, not in the $J_z=0$ cases;
each squared amplitude in Eq.~(7) has different top spin configuration 
factors according to the photon polarization.
Hence a combination of the $J_z=2$ and $J_z=0$ states
enables low scale quantum gravity
to modify the top spin configurations.
In fact finite mixing of two photon states is
inevitable from the practical point of view,
as shall be discussed below.
On the other hand, this mixing contaminates
the SM prediction of the suppression,
in a special top spin basis,
of 
$\gamma_R\gamma_L \to (t_\uparrow \bar{t}
_\uparrow ~{\rm or}~ t_\downarrow\bar{t}_\downarrow)$.

We briefly review the differential cross sections at practical
$\gamma\gamma$ colliders\cite{Ginzburg}.
From the head-on collisions between the laser
and energetic electron (or positron) beams,
high energy polarized photons are produced.
When $x$ denotes the fraction of the photon beam energy 
to initial electron beam energy, i.e., $x=E_\gamma/E$,
its maximum value is $x_{max}=z/(1+z)$
where $z=4E \omega_0/m_e^2$.
Here $E_\gamma$, $E$, $\omega_0$ are the photon, electron and laser beam 
energies, respectively. 
It is known that laser beam with too high energy would produce
$e^+e^-$ pair through collisions with the back-scattered photon beam,
reducing the $\gamma\gamma$ luminosity.
Thus the $z$ is optimized to be $2(1+\sqrt{2})$ which occurs
at the threshold for the electron pair production.
In the numerical analysis,
we consider the following cuts:
\begin{eqnarray}
&&-0.9 \le \cos \theta \le 0.9, \nonumber \\
&& \sqrt{0.4}\le x_{1(2)} \le x_{max}|_{z=2(1+\sqrt{2})}.
\end{eqnarray}

With given polarizations of the laser and parent electron beams 
the differential cross section 
is,
taking into account of their Compton back-scattering,
\begin{eqnarray}
\label{conv}
\frac{d\sigma}{d\cos\theta}&=&\frac{1}{32\pi s_{ee}}
\int \int dx_1dx_2 \frac{f(x_1)f(x_2)}{x_1x_2} \nonumber \\
&&\times \bigg[ \Big( \frac{1+\xi_2(x_1)\xi_2(x_2)}{2}\Big) 
\Big|{\cal M}_{J_z=0} \Big|^2 
+\Big( \frac{1-\xi_2(x_1)\xi_2(x_2)}{2}\Big) 
\Big|{\cal M}_{J_z=2} \Big|^2 \bigg],
\end{eqnarray}
where
\begin{eqnarray}
\label{sum}
\Big|{\cal M}_{J_z=0} \Big|^2 &=& \frac{1}{2}\bigg[
\Big|{\cal M}_{RR}\Big|^2 + \Big|{\cal M}_{LL}\Big|^2 \bigg], 
\\ \nonumber
\Big|{\cal M}_{J_z=2} \Big|^2 &=& \frac{1}{2}\bigg[
\Big|{\cal M}_{RL}\Big|^2 + \Big|{\cal M}_{LR}\Big|^2 \bigg]. 
\end{eqnarray}
And $f(x)$ and $\xi_2(x)$ are the photon number density
and the averaged circular polarization of the
back-scattered photon beams, respectively, both of which
of depend on the polarizations of the laser and electron beams
\cite{CCIS,Davoudiasl,choi}.
As $f(x)$ implies,
the back-scattered photons possess non-trivial energy spectrum,
which renders unavoidable finite mixing of the $J_z=2$ and $J_z=0$ states.
The $s_{ee}$ is the CM squared energy 
at $e^+ e^-$ collisions, related with that of
$\gamma\gamma$ collisions, $s$, by $s=x_1 x_2 s_{ee}$.
Since the $\gamma\gamma$ luminosity spectrum 
shows a narrow peak 
at around $x_{max}$
with an appropriate cut on the longitudinal momentum
of the back-scattered photons\cite{Ginzburg},
the approximations such as $t=x_1 x_2 t_{ee}$ and $u=x_1 x_2 u_{ee}$
are to be employed\cite{CCIS,Davoudiasl,choi}.


\section{Results}

We estimate various observables of the top quark pair
production at photon colliders.
Let us first observe that top spin information with $CP$ invariance 
permits only two independent cross sections,
\begin{equation}
\sigma_{\uparrow\uparrow} (= \sigma_{\downarrow\downarrow}) ,~~  
\sigma_{\uparrow\downarrow} (= \sigma_{\downarrow\uparrow}), 
\end{equation}
which can be easily seen from Eqs.~(\ref{cp}) and (\ref{sum}).
In Fig.~3, we present $\sigma_{\uparrow\uparrow}$ 
and $\sigma_{\uparrow\downarrow}$ with respect to
$\cos\xi$ at $\sqrt{s_{ee}}=1$ TeV
with unpolarized and/or polarized initial beams. 
The polarizations of the laser and electron beams
are allowed as 100\% and 90\%, respectively.
The $M_S$ is set to be 2.5 TeV. 
The convex curves denote the 
$(t_\uparrow \bar{t}_\downarrow)$ case, while
the concave ones do the $(t_\uparrow \bar{t}_\uparrow)$ case.
The solid lines describe the SM
results, and the dotted and dashed lines describe the results 
including low scale quantum gravity effects
with $\lambda=1$ and $\lambda=-1$, respectively.
Each value of $\cos\xi$ indicates different spin basis
of the top quark and anti-quark.

There are several interesting characteristics of 
photon colliders in producing top quark pair.
As can be first seen in Fig.~3, 
there is no special $\cos\xi$ at which
one top spin configuration dominates over the other, as discussed before.
The quantum gravity effects manifest themselves
as being factored out from the top spin configuration, 
i.e., independent of $\cos\xi$.
And irrespective of the polarizations of the laser and electron beams,
the effects are larger in the cases of $t_\uparrow \bar{t}_\downarrow$
than in those of $t_\uparrow \bar{t}_\uparrow$.
Equation (\ref{ampsq}) suggests its reason.
For instance, with the unpolarized beams,
the ratio $|{\mathcal M}_{RL}/{\mathcal M}_{RR}|^2$ hints 
the magnitude of the quantum gravity effects,
since the $J_z=2$ states possess the effects but the $J_z=0$ states do not.
Hence the top quark pair production channel with the 
$(t_\uparrow \bar{t}_\downarrow)$ is effective to probe
the quantum gravity effects.
Finally we observe that the results with the polarized
electron beams are of great interest and importance:
there exist two values of $\cos\xi$, i.e., 
two special spin bases where the equalities
$\sigma_{\uparrow \uparrow } = \sigma_{\uparrow \downarrow}$ hold true
as shown in Fig.~3 (c) and (d).
Considering a new observable, top spin asymmetry $A^{top}$,
\begin{equation}
A^{top} = {{ \sigma_{\uparrow\downarrow} -\sigma_{\uparrow\uparrow}}
\over{ \sigma_{\uparrow\downarrow} +\sigma_{\uparrow\uparrow}}},
\end{equation}
we suggest an optimal top spin 
basis defined by $\xi_0$ which satisfies
\begin{equation}
\sigma_{\uparrow\downarrow} ^{SM} ( \xi_0) =
\sigma_{\uparrow\uparrow} ^{SM} ( \xi_0)
\,.
\end{equation}
The definition guarantees that the observable
$A^{top}(\xi_0)$ vanishes in the SM
while retains non-vanishing values 
with the low scale quantum gravity effects.
The $A^{top}(\xi_0)$ as a function of $M_S$ is plotted
in Fig.~4.  
The solid lines indicate the SM predictions
at $2\sigma$ level,
while the dotted and dashed lines include
the quantum gravity effects of $\lambda=1$ and
$\lambda=-1$, respectively.
The case of $M_S=2.5$ TeV at $\sqrt{s_{ee}}=1.0$ TeV
causes about $A^{top}(\xi_0)$ of 6\%.

\vskip 0.5cm
\noindent
\begin{center}
\begin{tabular}{|c|cc|cc|cc|}
\hline
&
\multicolumn{2}{|c|}{$\sigma_{tot}$}
&
\multicolumn{2}{|c|}{$\sigma_{\uparrow\uparrow}$}
&
\multicolumn{2}{|c|}{$\sigma_{\uparrow\downarrow}$}
\\
$(P_{e1}, P_{e2}, P_{l1}, P_{l2})$ & ~$\lambda=1 $~ & ~$\lambda=-1 $~ &
~  $\lambda=1 $ & ~$\lambda=-1 $~ &
~  $\lambda=1 $~ &~ $\lambda=-1 $ ~\\
\hline
(0,0,0,0) & 4.1 & 4.1 & 2.6 & 2.8 & 4.3 & 4.3 \\
(0,0,1,1) & 4.0 & 4.0 & 2.5 & 2.7 & 4.2 & 4.2 \\
(0.9,0.9,0,0) & 3.2 & 3.2 & 1.7 & 2.0 & 3.7 & 3.7 \\
~(0.9,0.9,1,1)~~ & 3.2 & 3.2 & 1.8 & 2.0 & 3.5 & 3.6  \\
\hline
\end{tabular}
\end{center}
\vskip 0.3cm
{Table~I}. The photon collider bounds of $M_S$ in TeV at $2\sigma$
level according to the polarizations of the laser, the parent
electron beam, and top spin configurations.
\vskip 0.5cm

The $A^{top}(\xi_0)$ observables yield lower bounds
on the string scale $M_S$, which can be read off from Fig. 4.
In Table I, we summarize other lower bounds of the $M_S$,
which can be obtained by using various combination
of the polarizations of the laser and electron beams,
and by measuring the top spin configuration.
We have employed the luminosity ${\mathcal L} = 200 {\rm fb}^{-1} $.
As discussed before, the top spin state 
$(t_\uparrow \bar{t}_\downarrow)$  can constrain most strictly
the $M_S$, 
though the enhancement of the $M_S$ bound 
is not very significant.
This is mainly due to the statistical disadvantage of 
the top spin measurements.


\section{Conclusions}
The experimental feasibility of the determination of top 
spin configuration and construction of high energy linear 
photon colliders provides
a new window to probe physics beyond the standard model.
We have studied top pair production 
at $\gamma \gamma$ linear colliders
with the large extra dimension effects.
From the polarized scattering amplitudes in a general top spin basis,
it has been shown that only the $J_z=2$ states receive the effects,
which do not affect the top spin configuration.
The non-trivial energy spectrum of the Compton back-scattered photons
leads to finite mixing of the $J_z=0$ and $J_z=2$ states,
as well as permitting low scale quantum gravity to
play a role in modifying the top spin configuration.
Considering the Compton scattering also,
total cross sections in a general top spin basis
have been presented; the large extra dimension effects
are shown to be larger for the $(t_\uparrow\bar{t}_\downarrow)$
configuration than for the $(t_\uparrow\bar{t}_\uparrow)$ one.
On the other hand, this practical mixing 
of the $J_z=0$ and $J_z=2$ states
is shown to eliminate the presence of a top spin basis 
in which one of the final
spin configurations is dominant and the other suppressed.
Instead we have noticed that a special spin basis,
especially with the polarized parent electron beams,
exist such that $\sigma^{SM}(t_\uparrow\bar{t}_\downarrow)
=\sigma^{SM}(t_\uparrow\bar{t}_\uparrow)$ in the SM.
Accordingly introduced is the top spin asymmetry $A^{top}$
of which the SM predictions vanish in this optimal top spin basis,
which can be one of the most effective tools.
Low scale quantum gravity 
at $\sqrt{s_{ee}}=1$ TeV
with the string scale $M_S=2.5$ TeV
predicts about 6$\%$ of $A^{top}$ in the optimal top spin basis.
From various observables,
we have also estimated the lower bounds of the $M_S$
in the range of 1.7 to 4.3 TeV, summarized in Table I. 
The measurements of the top spin configuration 
as $(t_{\uparrow} \bar{t}_{\downarrow})$
improve the lower bounds for $M_S$ more than 200 GeV.

\acknowledgements

This work is supported 
in part by the Korean Science and Engineering Foundation (KOSEF) 
through the Center for Theoretical Physics (CTP)
at Seoul National University.

\newpage
\begin{center}
{\large \bf Figure Captions}
\end{center}
\vskip 2cm

\begin{description}

\item
Fig.~1 :
Feynman diagrams for the $\gamma \gamma  \rightarrow t \bar{t}$ process
with the large extra dimension effects.

\item
Fig.~2 :
The generic spin basis: (a) in the top quark rest frame; (b) in the top
anti-quark rest frame.

\item
Fig.~3 :
Total cross sections as a function of $\cos \xi$: (a) with the 
unpolarized beams, i.e., $(P_{e1}, P_{e2},P_{l1},P_{l2})=(0,0,0,0)$;
(b) $(0,0,1,1)$; (c) $(0.9,0.9,0,0)$; 
(d) $(0.9,0.9,1,1)$.
The convex curves denote the
$(t_\uparrow \bar{t}_\downarrow)$ case, while
the concave ones do the $(t_\uparrow \bar{t}_\uparrow)$ case.
The solid lines describe the SM
results, and the dotted and dashed lines describe the results
including low scale quantum gravity effects
with $\lambda=1$ and $\lambda=-1$, respectively.

\item
Fig.~4 :
The optimal top spin asymmetries with respect to $M_S$ 
with unpolarized and/or polarized initial beams.
The solid lines denote $2\,\sigma$ level predictions of the SM.
The dotted and dashed lines incorporate low scale quantum gravity effects 
in the cases of $\lambda = 1$ and $\lambda = -1$, respectively.

\end{description}

\newpage

\begin{figure}[h]
\vspace{7.0cm}
\centerline{\epsfig{file=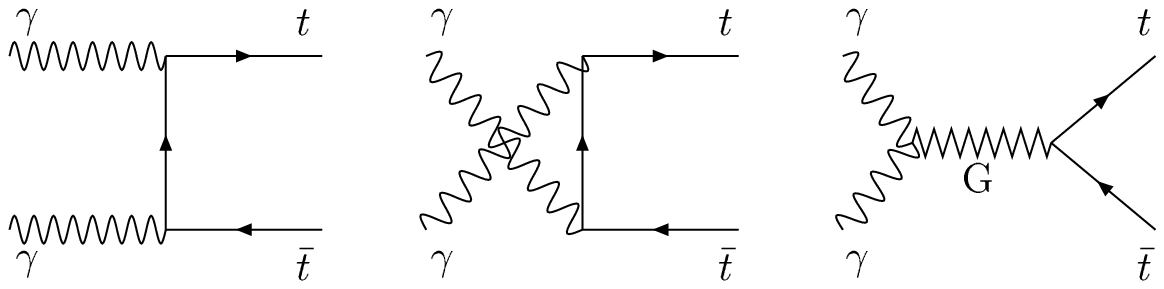}}
\end{figure}
\vspace{2.0cm}
\begin{center}
{\large Fig. 1}
\end{center}

\newpage
~
\vspace{5cm}
\begin{figure}[th]
\centering
\centerline{\epsfig{file=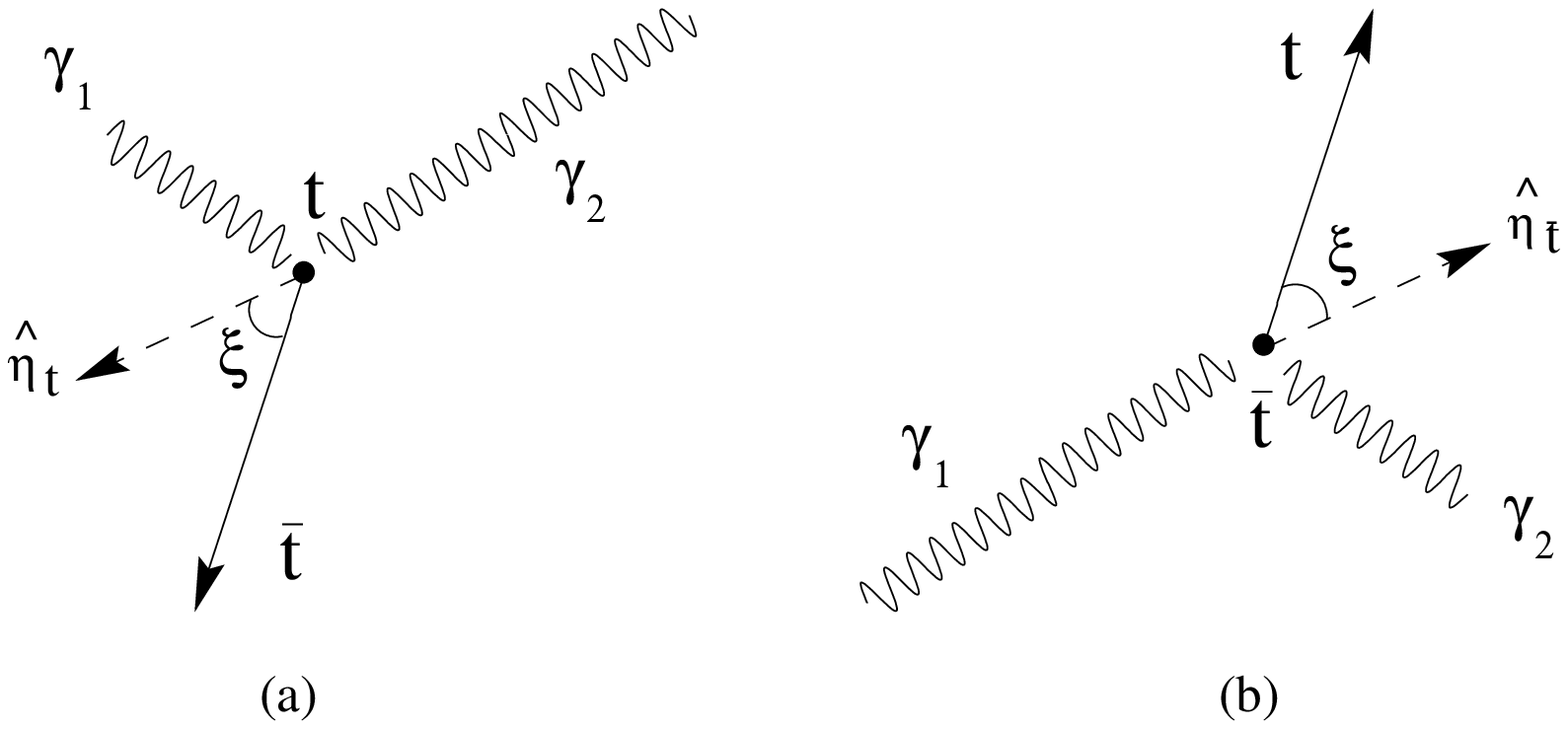}}
\end{figure}
\vspace{1.0cm}
\begin{center}
{\large Fig. 2}
\end{center}

\newpage
~
\vspace{1cm}
\begin{figure}[th]
\centering
\centerline{\epsfig{file=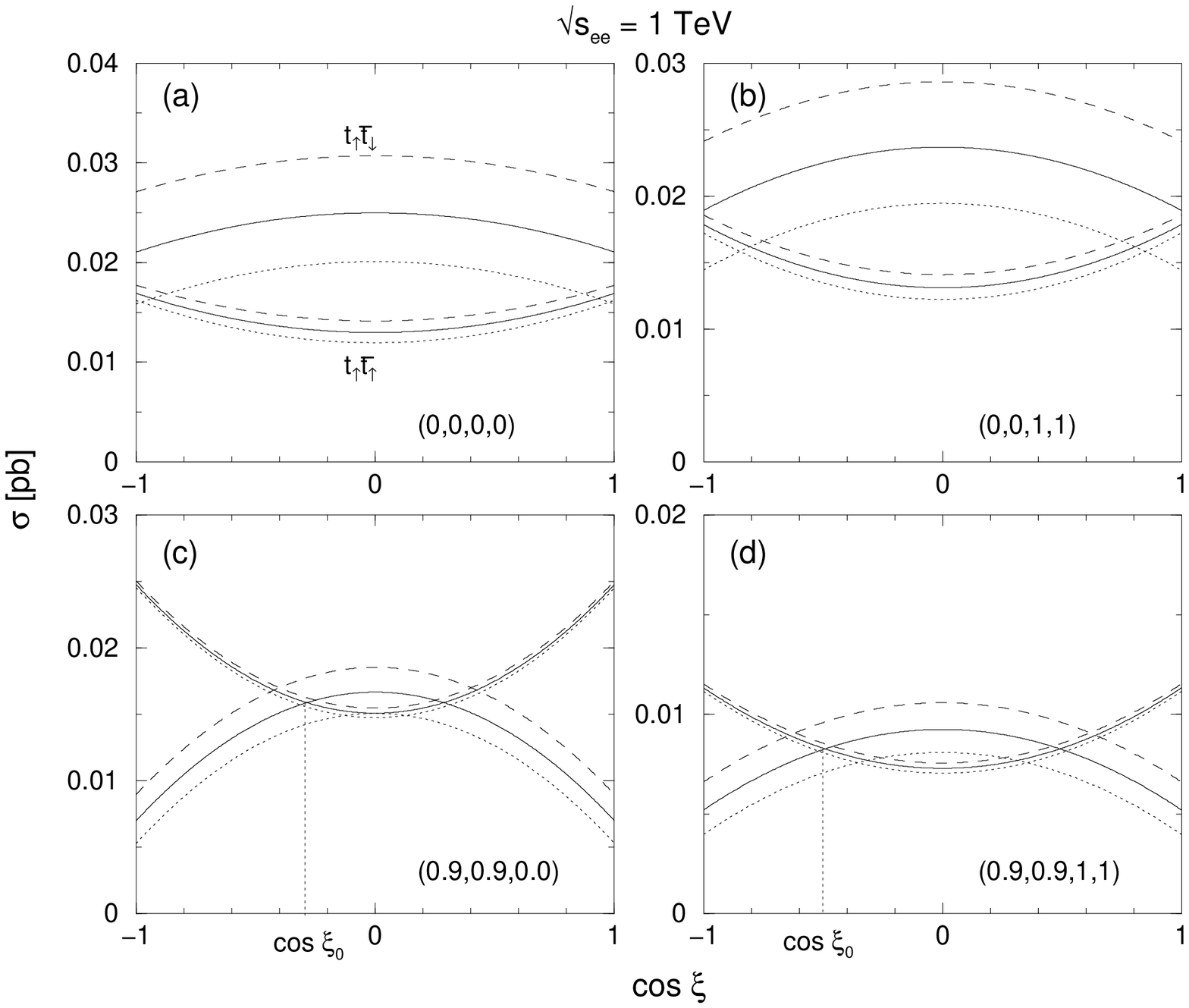}}
\end{figure}
\vspace{1.0cm}
\begin{center}
{\large Fig. 3}
\end{center}

\newpage
~
\vspace{2.5cm}
\begin{figure}[th]
\centering
\centerline{\epsfig{file=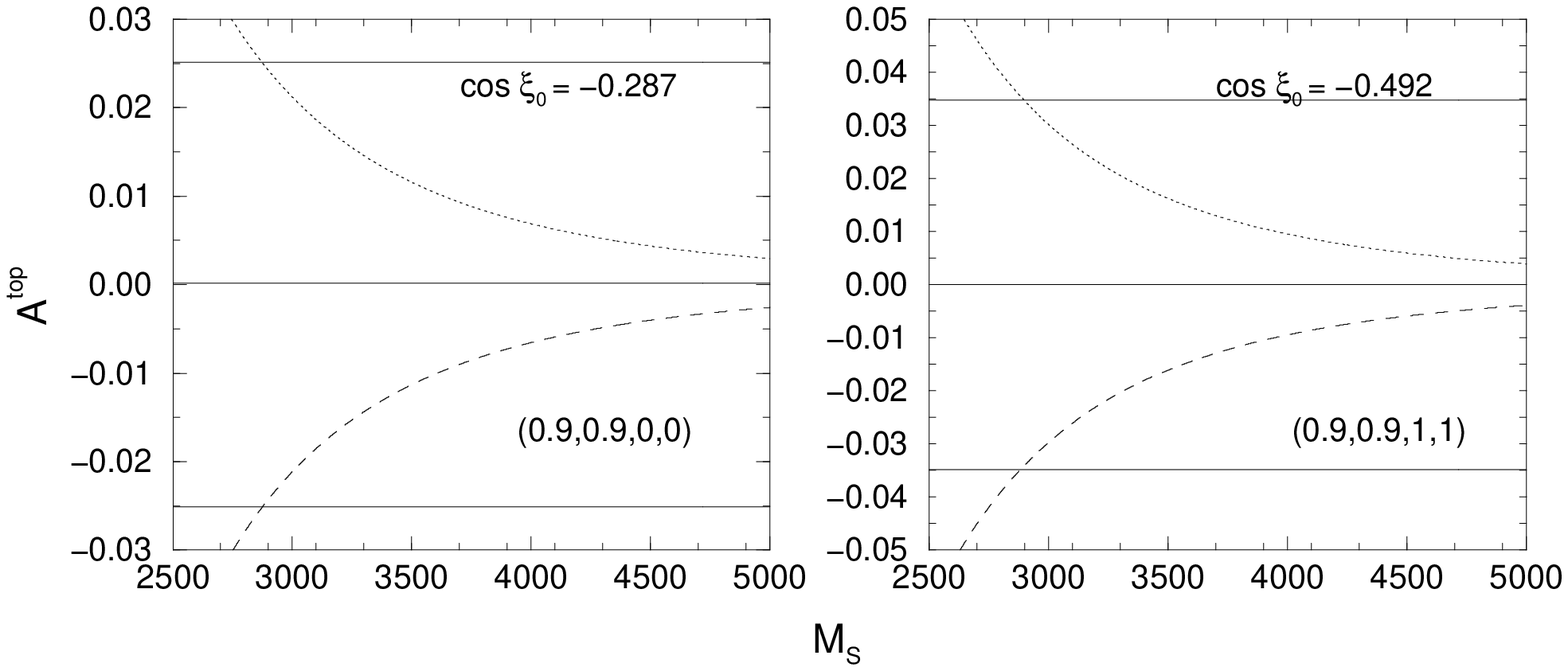}}
\end{figure}
\vspace{1.0cm}
\begin{center}
{\large Fig. 4}
\end{center}

\end{document}